\documentstyle[prl,multicol,aps]{revtex}
\input{epsf}
\tighten

\begin{document}
\preprint{LA-UR-96-????}

\draft

\title{Pairing Correlations in the Two-Dimensional Hubbard Model}

\author{Shiwei Zhang}
\address{Department of Physics, The Ohio State University, Columbus, OH 43210\\
Department of Applied Science 
and Department of Physics, College of William and Mary,
Williamsburg, VA 23187 \ \cite{byline}}

\author{J.~Carlson and J.E.~Gubernatis}
\address{Theoretical Division,
Los Alamos National Laboratory, Los Alamos, NM 87545}

\date{\today}
\maketitle

\begin{abstract}
We present the results of a quantum Monte Carlo study of the extended
$s$ and the $d_{x^2-y^2}$ pairing correlation functions for the
two-dimensional Hubbard model, computed with the constrained-path
method. For small lattice sizes and weak interactions, we find that
the $d_{x^2-y^2}$ pairing correlations are stronger than the extended
$s$ pairing correlations and are positive when the pair separation
exceeds several lattice constants. As the system size or the
interaction strength increases, the magnitude of the long-range part
of both correlation functions vanishes.

\end{abstract}

\bigskip


\begin{multicols}{2}
\narrowtext

Since the discovery of high-temperature superconductivity, the
two-dimensional Hubbard model has been the subject of an
unprecedented level of theoretical activity to discover whether it can
serve as the paradigm for this novel and important
phenomenon. Particularly with respect to magnetic properties
\cite{kampf94}, the physics of the model qualitatively represents the
behavior of the real materials.  For example, at half-filling the
model is an anti-ferromagnetic insulator. Upon doping the
anti-ferromagnetism rapidly becomes strongly suppressed. This behavior
is observed in the cuprate superconductors. A variety of calculations
also predict\cite{scalapino95} that the doped model
exhibits an attractive interaction between pairs; the 
$s$  and $d_{x^2-y^2}$ symmetries of this attraction are consistent with the
likely symmetries of the experimentally measured superconducting gap
\cite{scalapino95}. Yet unobserved, however, is convincing evidence
that the attractive interaction leads to a ground state with
off-diagonal long-range order \cite{kampf94,scalapino95}.  In this
paper we will present results from quantum Monte Carlo (QMC)
simulations which suggest that the long-range extended $s$ and $d_{x^2-y^2}$
pairing correlations in fact vanish in the thermodynamic limit.

The fundamental difficulty in deciding whether the two-dimensional
Hubbard model superconducts is the absence of an exact
solution. Approximate solutions have often been uncontrolled,
difficult to benchmark, and conflicting.  On several key points,
computer simulations have provided important information.  In fact,
the possible existence of superconductivity in the Hubbard model was
suggested by the results of a QMC simulation
before the discovery of the high temperature superconducting materials
\cite{hirsch85}.

Numerical approaches have, however, had their own difficulties,
typically being limited to small system sizes, high temperatures, and
selected electron fillings. Quantum Monte Carlo methods, for example,
experience the infamous fermion sign problem \cite{schmidt84,loh90},
which causes an exponential growth in the variance of the computed
results and hence an exponential growth in computer time as the
lattice size is increased and the temperature is lowered.  While QMC
simulations have shown indications of pairing correlations,
uncertainty has remained because of their restriction to relatively
small lattice sizes and high temperatures \cite{scalapino95,moreo91}.

Here, using our new constrained path Monte Carlo (CPMC) method
\cite{zhang95}, we discuss the behaviors of the extended $s$ and
$d_{x^2-y^2}$ pairing correlations obtained from simulations free of
such restrictions. Contrary to standard algorithms
\cite{white89,loh92}, 
this new ground-state ($T=0$\,K) method 
exhibits {\it algebraic\/} scaling of computer time with system
size. It eliminates the exponential growth of variances by 
what we call the constrained-path approximation. In a variety of 
benchmarking calculations \cite{zhang95}, the CPMC method has yielded accurate 
estimates of the energy as well as other ground-state observables. 

We considered the following familiar form of
the two-dimensional Hubbard model on a square lattice 
of $N=L\times L$ sites with $N_\sigma$ ( $\sigma=\uparrow,\downarrow$)
electrons:
\begin{equation}
  H =-t\sum_{\langle ij \rangle \sigma} (c_{i \sigma}^\dagger
    c_{j\sigma} + c_{j \sigma}^\dagger c_{i\sigma}) + U \sum_i n_{i
    \uparrow} n_{i \downarrow}.
\label{eq:H}
\end{equation}
we took $t=1$ and assumed periodic boundary conditions.
The pairing correlation function we computed is 
\begin{equation}
P_\alpha({ l}) = \langle \Delta^\dagger_\alpha({ l }) \Delta_\alpha({ 0})\rangle,
\label{eq:pairing_def}
\end{equation}
where $\alpha$ indicates the nature of pairing. The pair field-operator 
at site $l$ is 
$\Delta_\alpha ({ l}) = 
\sum_\delta f_\alpha(\delta) [c_{l\uparrow} c_{l+\delta\,\downarrow}
-c_{l\downarrow} c_{l+\delta\,\uparrow}]$, 
 where ${ \delta}$
is $(\pm 1,0)$ and $(0,\pm 1)$. For extended $s$ pairing, $f_{s} (
\delta )=1$. For $d_{x^2-y^2}$ pairing, $ f_d (\delta)$ is $1$
when ${ \delta}=(\pm 1,0)$ and $-1$ otherwise.

To facilitate contact with prior simulations, we also examined the
``vertex contribution'' to the 
correlation function \cite{white89} defined by
\begin{equation}
   \bar P_\alpha ({ l}) = P_\alpha(l)
               - \langle \Delta^\dagger_\alpha({ l})
                 \Delta_\alpha({ 0})\rangle_0
\label{eq:pairing_vertex}
\end{equation}
The second function on the right is shorthand notation for the
uncorrelated pairing correlation. For each term in $P_\alpha(l)$
like $\langle c_\uparrow^\dagger c_\uparrow\,c_\downarrow^\dagger
c_\downarrow \rangle$, it has a term like
$\langle c_\uparrow^\dagger c_\uparrow \rangle
\langle c_\downarrow^\dagger c_\downarrow \rangle$.

Our numerical method is extensively described and benchmarked
elsewhere \cite{zhang95}. Here we only discuss its basic
approximation.  In this method, the ground-state wave function
$|\Psi_0\rangle$ is projected from a known initial wave function
$|\Psi_T\rangle$ by a branching random walk in an over-complete space
of Slater determinants $|\phi\rangle$.  In such a space, we can write
$|\Psi_0\rangle = \sum_\phi \chi(\phi) |\phi\rangle$, where
$\chi(\phi)>0$.  The random walk produces an ensemble of
$|\phi\rangle$, called random walkers, which represent
$|\Psi_0\rangle$ in the sense that their distribution is a Monte Carlo
sampling of $\chi(\phi)$.

To completely specify the ground-state wave function, 
only determinants satisfying $\langle\Psi_0|\phi\rangle>0$
are needed; hence, $|\Psi_0\rangle$ resides in either of two degenerate 
halves of the Slater determinant space,
separated by a nodal plane ${\cal N}$ that is defined by
$\langle\Psi_0|\phi\rangle=0$. The sign problem occurs because
walkers can cross ${\cal N}$ as their orbitals 
evolve continuously in the random walk. Asymptotically 
they populate the two halves equally, leading to
an ensemble that has zero overlap with $|\Psi_0\rangle$.
If ${\cal N}$ were known, we would simply constrain the random walk 
to one half of the space and obtain
an exact solution of Schr\"odinger's equation.
Without {\it a priori\/} knowledge of ${\cal N}$, we use a
trial wave function $|\Psi_T\rangle$ 
and require $\langle\Psi_T|\phi\rangle>0$.  
The random walk again solves Schr\"odinger's equation
in determinant space, but 
under an approximate boundary-condition. 
This is what we call the constrained-path approximation. 

The ground-state energy computed by the CPMC method is an upper bound. 
The quality of the calculation clearly depends on
the trial wave function $|\Psi_T\rangle$. Since the constraint 
only involves the overall sign of its overlap with any determinant
$|\phi\rangle$, it seems reasonable to expect the results to show
insensitivity to $|\Psi_T\rangle$.
Through extensive benchmarking on the Hubbard model,
we have found that simple choices of this function can give very good results
\cite{zhang95}.  In the calculations
reported here we took $|\Psi_T\rangle$ to be a
single Slater determinant.  For closed-shell electron fillings,
we used the free-electron ($U=0$) wave
function. For open-shell fillings, we used
unrestricted Hartree-Fock (uHF) solutions. For the latter, we have found
that uHF solutions obtained with low $U$ values ($<1$), i.e., 
those resembling free-electron wave functions, tend to be good
choices for $|\Psi_T\rangle$ for $U$ up to $ 8$.

As a calibration of our method,  we compare in Table~1 its prediction
for the $d_{x^2-y^2}$ pairing correlation function to that obtained by
an exact diagonalization calculation \cite{moreo_private} of a
$4\times 4$ lattice with a closed shell filling of
$N_\uparrow=N_\downarrow=5$ and with $U=2$ and 4. At each 
location of the pairs, we reproduce the exact result within an
error of 1 \% or less. (Here, in order to compare with the
exact diagonalization data we computed $\langle \Delta_d({ l })
\Delta^\dagger_d({ 0})\rangle$ with $\Delta_d(l)=
   c_{l\uparrow}\sum_{ \delta} f_d(\delta) c_{l+\delta\,\downarrow}$.)

As a further calibration of our method, we show in Fig.~1 the
long-range portion of the $d_{x^2-y^2}$ pairing correlation $P_d(l)$
as a function of $|l|$ for a half-filled $8\times 8$ system at
$U=4$. At half-filling, the standard auxiliary-field quantum Monte
Carlo (AFQMC) method\cite{loh92} has no sign problem and is exact, and
the CPMC method can be made exact by removing the constrained path
condition. In this CPMC calculation, however, we deliberately kept the
constraint and used for $|\Psi_T\rangle$ the uHF solution of the system
at $U=0.5$. In the figure, the computed $P_d(l)$ from the CPMC simulation is
compared with exact AFQMC results. Also shown is the result predicted
by $|\Psi_T\rangle$.
The inset shows, as a function of electron filling
$(N_\uparrow+N_\downarrow)/N$, the relative difference between the
ground-state energies calculated by CPMC and AFQMC simulations.  The
point indicated by the arrow corresponds to the CPMC calculation shown
in the main graph.  At $1\%$, this difference represents the
largest systematic error in the CPMC calculation of the energy. With
the energy as a gauge, the CPMC calculation at 1/2-filling would
appear to be of the poorest quality; yet, we see that it still yields
an accurate $P_d(l)$.  The magnitude and range of these
correlations is comparable to those we now discuss for the doped
cases.

Figure~2 shows the long-range part of $P_d(l)$ as a function of $|l|$
for a $12\times 12$ lattice at $U=2$, 4, and 8.  Here, the electron
filling is 0.85, which corresponds to a closed shell case with
$N_\uparrow=N_\downarrow=61$.  Figure~2a, the $U=2$ case, shows three
different evaluations of this correlation function. One is the
free-electron prediction for the pairing.  Another is based on
definition (\ref{eq:pairing_def}), and the third is the vertex
contribution to this definition. Figure~2b shows the same set of
curves for $U=4$ while the inset to Figure~2b shows the $U=8$ results
with the vertex contribution omitted for clarity.  These three sets of
curves show that $P_d(l)$ is smaller at all three vaules of $U$ than
the non-interacting case. They also show that increasing $U$ causes
the long-range correlations, including the vertex contribution, to
vanish. At $U=8$, despite the large error bars, the correlations are
reduced to simply fluctuating around zero. We also see that the vertex
contribution is a fairly flat function of pair separation up to
$U=4$. This flat region is the ``plateau'' observed in
Ref.~\cite{husselin96} for calculations up to $U=2$.  In this work the
Hubbard model was studied with next near-neighbor hopping and the
existence of the ``plateau'' was attributed to its presence. Our results
show that the ``plateau'' behavior is no less pronounced in the simple
Hubbard model. As $U$ increases to $8$, however, it vanishes as
$P_d(l)$ does.

In Fig.~3, we address the question of what happens to these
``long-range'' correlations if the lattice size is increased to
$16\times 16$. Here, for a closed shell case with the same electron
filling of 0.85 ($N_\uparrow=N_\downarrow=109$), we show the CPMC
results for $U=2$ and $4$. First, we notice that as in Fig.~2
increasing interaction strength eventually causes the correlations to
vanish but now they vanish by $U=4$. The $U=4$ case is shown in
Fig.~3b; the accuracy is still sufficient to discern the
irregular oscillations of $P_d(l)$ around zero. In Fig.~3a the vertex
contribution is again relatively flat, but nearly zero. Compared to
Fig.~2a, it has decreased with the increase in lattice size.  At $U=4$
it has in fact vanished and it not shown for clarity. We note that
we have also carried out calculations with a second neighbor hopping
and did not find any qualitatively different behavior.

A representative result for the extended $s$ pairing correlation
function $P_s(l)$ is shown in Fig.~4 for the same system as in Fig.~2.
The pairing correlation function is shown for the whole range of
$|l|$.  Its short-distance magnitude is much greater than that in the
tail. We mention that the $d_{x^2-y^2}$-wave pairing correlation shows
the same general behavior. In both cases, the short-range correlation
actually increases as $U$ is increased from zero. Hence, the often
used integrated pairing correlation function, or equivalently the
${\bf k} = (0,0)$-component of $P_\alpha$ in momentum space, is {\it
not\/} a good indicator of superconductivity. Comparing the inset with
Fig.~2b, we see that the long-range extended $s$-wave pairing is at
least an order of magnitude weaker than the $d_{x^2-y^2}$
pairing. Indeed it is already fluctuating around zero.

We also studied pairing correlation functions for other electron
fillings, lattice sizes, and interaction strengths. These results
re-enforce those represented above and will be reported
elsewhere. For example, at open-shell fillings, even though the
calculations experience significant increases in variances due to a
poorer $|\Psi_T\rangle$'s, results through $12\times 12$
systems do not appear to show any significant changes compared to the
closed-shell ones.

We reiterate that, due to the constrained-path approximation, the
correlation functions computed here are approximate. While the
systematic error appears small when compared with exact
diagonalization and exact QMC results, we cannot exclude the
possibility that as the lattice size increases our systematic error
increases and an underestimation of the correlations develops.  By the
same token we cannot exclude overestimation. However, results like
those in Table~1 and Fig.~1, plus a variety of other benchmarks
\cite{zhang95}, indicate that our systematic error is
typically small. In fact, it is often orders of magnitude
smaller than the statistical error in simulations using the standard
AFQMC method. 

With this very small statistical error, we have pulled pairing
correlations ``out of the noise,'' and have shown examples that for a
given system size they disappear as the interaction strength increases
and for a given interaction strength they disappear as the system size
increases. We note that similar behavior exists for the
non-interacting problem and the half-filled case.

We have also computed the lattice size, interaction strength, and
electron filling dependence of the ground state energy, electron
momentum distribution, and static spin-spin correlation function.  We
will report these results elsewhere.

We thank A.~Moreo for the exact diagonalization results.  We
acknowledge helpful conversations with D.~L.~Cox, D.~Pines,
S.~Trugman, and J.~W.~Wilkins. Our calculations were performed on the
SP2 computer at the Cornell Theory Center.  We
are very grateful for this support. The work was supported by the
Department of Energy.

\newpage


%
%

\begin{figure}
\epsfxsize=2.8in
\epsfysize=3.0in
\centerline{\epsfbox{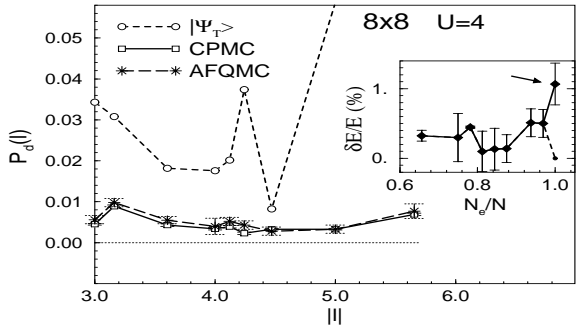}}
\vskip-1.35in
\caption{Long-range behavior of the $d_{x^2-y^2}$ pairing correlation
function versus distance for a half-filled $8\times 8$ lattice at
$U=4$ computed with $|\Psi_T\rangle$ and by the CPMC and AFQMC
methods. The inset shows the relative difference between the CPMC and
AFQMC energies as a function of electron filling. The error bars
are statistical in origin and mainly associated with the AFQMC
results.}

\label{fig1}
\end{figure}

\begin{figure}
\vspace{0.05in}
\epsfxsize=2.7in
\epsfysize=3.4in
\centerline{\epsfbox{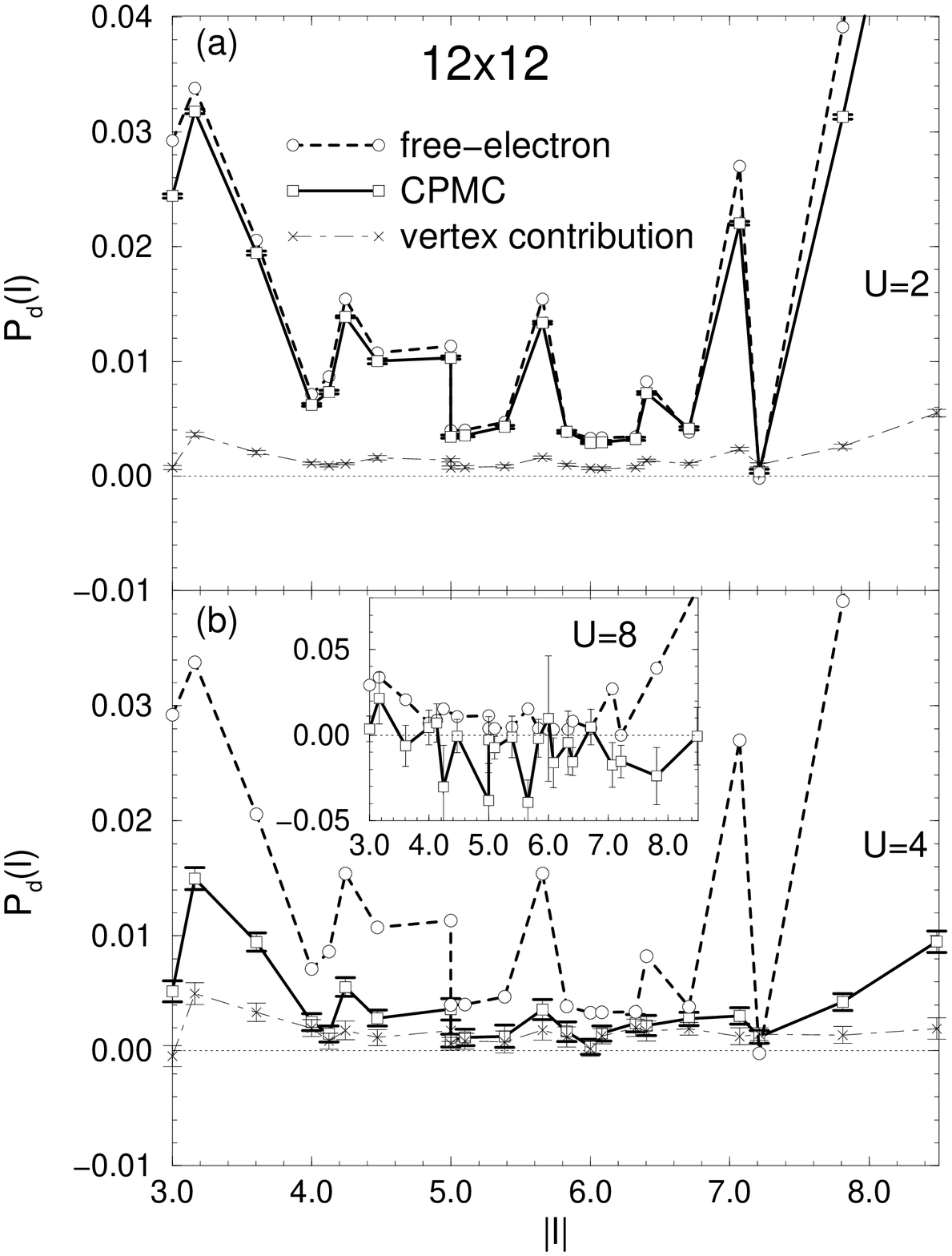}}
\vspace{0.2in}
\caption{Long-range behavior of the $d_{x^2-y^2}$ pairing correlation
function versus distance for 0.85 filled $12\times 12$ lattice at
$U=2$, 4, and 8. This behavior is shown for the free-electron
and CPMC calculations. Also shown is the vertex contribution.}
\label{fig2}
\end{figure}

\begin{figure}
\epsfxsize=2.7in
\epsfysize=3.4in
\centerline{\epsfbox{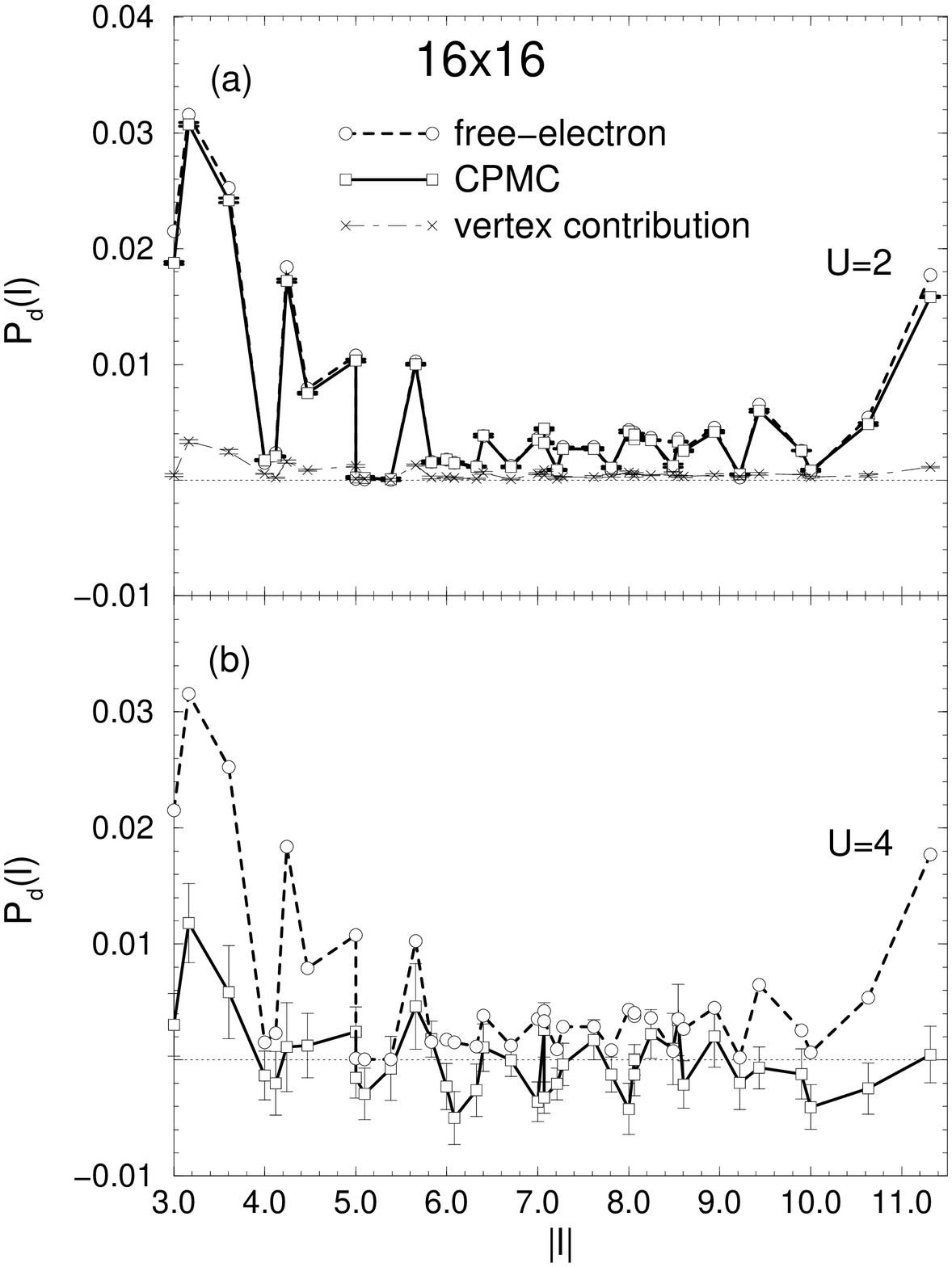}}
\vspace{0.2in}
\caption{Long-range behavior of the $d_{x^2-y^2}$ pairing correlation
function versus distance for a 0.85 filled $16\times 16$ lattice at
$U=2$ and 4.  This behavior is shown for the free-electron
and CPMC calculations. Also shown is the vertex contribution.}
\label{fig3}
\end{figure}

\begin{figure}
\vspace{0.1in}
\epsfxsize=2.8in
\epsfysize=3.0in
\centerline{\epsfbox{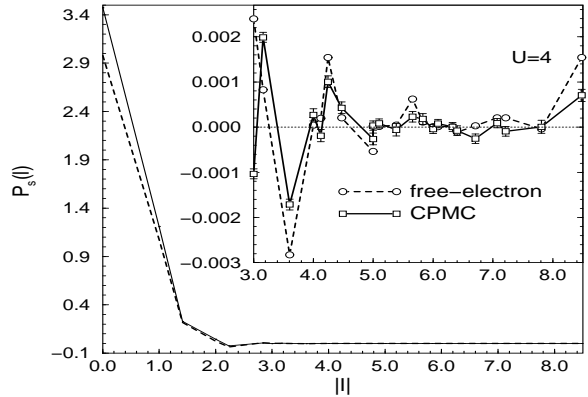}}
\vspace{0.1in}
\vskip-1.0in
\caption{Extended $s$ pairing correlation function versus distance for a
$12\times 12$ lattice at $U=4$ with an electron filling of 0.85. The
inset shows long-range behavior of the free-electron case
compared with the CPMC result.}
\label{fig4}
\end{figure}

\end{multicols}

%
%
%

\widetext

\begin{table}
\caption{Comparison of the CPMC $d_{x^2-y^2}$ pairing correlation
function with exact
  diagonalization results \protect\cite{moreo_private} as a function of
  pair separation $l=(l_x,l_y)$.  The system   size is   $4\times 4$
  with $N_\uparrow  
  =N_\downarrow =5$.  In the CPMC calculations the free-electron wave
  function  was used for   $|\Psi_T\rangle$. Statistical errors in these
  calculations are in the last digit and are indicated in parentheses.}
\label{tab1}
\begin{tabular}{c@{\ \ \ \ }l r@{}l r@{}l r@{}l r@{}l r@{}l r@{}l}
$U$ &       & \multicolumn{2}{c}{$(0,0)$}     
            & \multicolumn{2}{c}{$(1,0)$}
            & \multicolumn{2}{c}{$(2,0)$}
            & \multicolumn{2}{c}{$(1,1)$}
            & \multicolumn{2}{c}{$(2,1)$}
            & \multicolumn{2}{c}{$(2,2)$} \\ \tableline
$2$ & CPMC  & 2.&0672(2) & 0.&0924(1) & $-$0.&1121(1) & 0.&1140(1)
                  & 0.&0284(1)   & 0.&1779(2) \\
    & exact & 2.&06693   & 0.&09223   & $-$0.&11187    & 0.&11381 
                  & 0.&02840     & 0.&17793   \\ \hline
$4$ & CPMC  & 2.&0635(5) & 0.&0876(3) & $-$0.&0941(3) & 0.&1006(4)
                  & 0.&0246(2)   & 0.&1532(6) \\
    & exact & 2.&06345   & 0.&08714   & $-$0.&09422    & 0.&10013
                  & 0.&02453     & 0.&15302   \\
\end{tabular}
\end{table}

\end{document}